# The calibration and electron energy reconstruction of the BGO ECAL of the DAMPE detector

Zhiyong Zhang,[1] Chi Wang,[1] Jianing Dong,[1] Yifeng Wei,[1] Sicheng Wen,[1,2] Yunlong Zhang,[*,1] Zhiying Li,[1] Changqing Feng,[1] Shanshan Gao,[1] ZhongTao Shen,[1] Deliang Zhang,[1] Junbin Zhang,[1] Qi Wang,[1] SiYuan Ma,[1] Di Yang,[1] Di Jiang,[1] Dengyi Chen,[2] Yiming Hu,[2] Guangshun Huang,[1] Xiaolian Wang,[1] Zizong Xu,[1] Shubin Liu,[1] Qi An,[1] Yizhong Gong.[2]

[1]*State Key Laboratory of Particle Detection and Electronics (IHEP-USTC), University of Science and Technology of China, Hefei, 230026, China*

[2]*Purple Mountain Observatory, Chinese Academy of Sciences, Nanjing, 210000, China*

*Abstract:* The DArk Matter Particle Explorer (DAMPE) is a space experiment designed to search for dark matter indirectly by measuring the spectra of photons, electrons, and positrons up to 10 TeV. The BGO electromagnetic calorimeter (ECAL) is its main sub-detector for energy measurement. In this paper, the instrumentation and development of the BGO ECAL is briefly described. The calibration on the ground, including the pedestal, minimum ionizing particle (MIP) peak, dynode ratio, and attenuation length with the cosmic rays and beam particles is discussed in detail. Also, the energy reconstruction results of the electrons from the beam test are presented.

## 1. Introduction

Weakly interacting massive particles (WIMPs) are historically the most popular dark matter candidates [1].So far, there are many direct and indirect detection experiments [2-5] currently running or planned, aiming to search for evidence of them. The indirect detection experiments search for the products of WIMP annihilation or decay [6-7].The DAMPE detector is designed to search for dark matter indirectly by measuring the spectra of photons, electrons, and positrons with a wide energy range from 5 GeV to 10 TeV and a good energy resolution of 1.5% at 800 GeV in space [8-9]. The DAMPE detector was launched in December of 2015 and is operated at an altitude of approximate 500-kilometer, on a sun synchronic satellite orbit around the earth, stably oriented to the zenith.

The DAMPE detector is composed of four sub-detectors (Fig. 1). First, the plastic scintillator detector array (PSD), which consists of two layers (X-Y) of scintillator strips, is used to discriminate among heavy ion species. It is also used to identify electrons and gamma rays. Second, the silicon-Tungsten tracker (STK) has six planes (each plane has two orthogonal layers) of silicon micro-strip trackers and three layers of Tungsten plates of 1.0mm thickness (black active lines in Fig. 1 STK) inserted in front of tracking layers 2, 3, and 4 for photon conversion. It is designed to provide tracking and $e^{\pm}/\gamma$ identification. Third, there is the BGO electromagnetic calorimeter (BGO ECAL), which is the focus of this paper, and most of the details regarding it will be presented in the following section. Fourth, the neutron detector (NUD) at the bottom of the DAMPE aims to improve e/p identification capacity by detecting the thermal neutrons produced in the BGO ECAL by high-energy protons [10].

As a satellite-based experiment, the DAMPE detector required much consideration with regard to design and testing in order to be operated in space. Some important points, such as its weight, power, temperature control, mechanical structure, etc., were taken into account throughout the design process. Many studies on these elements have already been carried out. For instance, the temperature effect, light yield, uniformity of the light yield, and the attenuation coefficient were studied for the crystals and aging effect, operation in a vacuum, charge ratios between dynodes, etc., were studied for the PMTs. After the construction of the sub-detectors, many experiments simulating the rigorous environment in orbit were employed for each sub-detector individually to make sure that they could be operated well in space. The tests included electromagnetic compatibility, which is used to check the interference rejection capability and estimate the interferences to the satellite. The vibration test studies the mechanical properties that should withstand the G-forces for launching. The thermal cycle test checks stability and reliability with in a large temperature range from -20 to 40 degrees centigrade. And the thermal vacuum test simulates the environment of space and temperature variations. Following construction of the entire DAMPE detector, the same environmental tests were required again before launch.

*E-mail: ylzhang@ustc.edu.cn

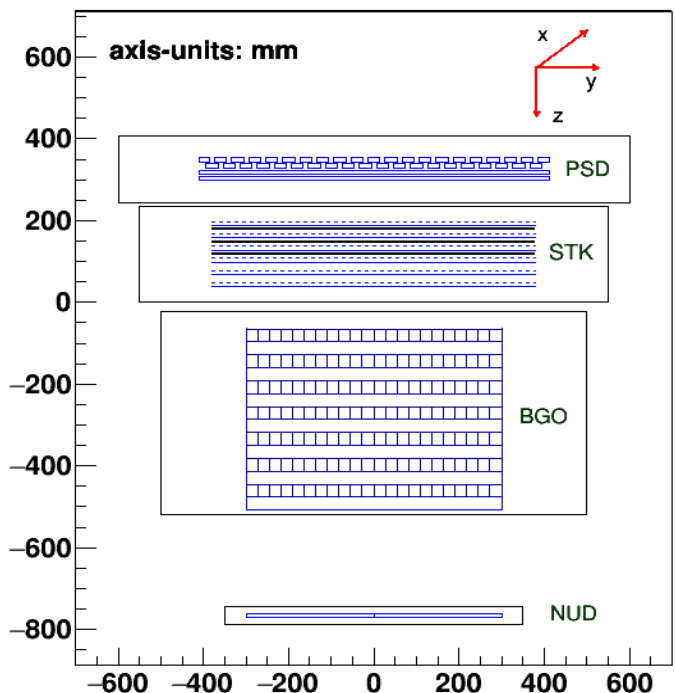

Fig.1. The structure diagram of the DAMPE detector. The blue box and lines show the sensitive detectors. There the y axis points to the sun, and -z is oriented to the zenith in the red coordinates.

## 2.The BGO Calorimeter

### 2.1 Detector

The BGO ECAL is the main sub-detector for energy measurement, so it is designed to cover a large energy range from 5GeV to 10TeV with a good energy resolution of 1.5% at 800GeV. The primary purposes of the BGO ECAL are to measure the energy deposition due to the particle shower that is produced by the $e^{\pm}$, γ and image their shower development profile, thereby providing an important hadron discriminator. Therefore, it is designed to contain fourteen layers of BGO crystals, about 31radiation lengths. Each layer is composed of 22 BGO bars in dimensions of 25× 25 × 600 mm³, with a PMT coupled on each end of the BGO crystal to collect scintillation light from the bar. Each single crystal and its 2 PMTs constitute a minimum detection unit (MDU, Fig.2) of the BGO ECAL. The layers of BGO bars are alternated in an orthogonal way to measure the deposited energy and shape of the nuclear and electromagnetic showers developed in the BGO ECAL.

### 2.2Minimum detection unit

The energy deposit from a MIP passing through a 2.5 cm-thick BGO bar is about 23 MeV. A Geant4-based simulation indicates that the maximum energy deposit in such a BGO bar is about 2 TeV from a 10

*E-mail: ylzhang@ustc.edu.cn

TeV electron, corresponding to about $10^5$ MIPs. A minimum measure able energy deposit of 0.5 MIPs is required for shower shape reconstruction with reasonable precision demanded by particle identification. Thus, each MDU should cover energy measurements in a range from 0.5 MIPs to $10^5$ MIPs, corresponding to a high dynamic range of $2 \times 10^5$. In order to cover such a large dynamic range of energy measurements, a multi-dynode readout PMT base board was conceived and designed, by which the signals are readouts from the different sensitive dynodes 2, 5, and 8 (Dy2, Dy5, and Dy8) [11].

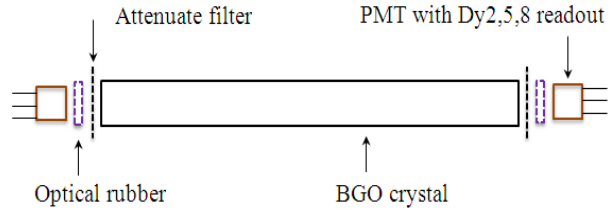

Fig.2.The minimum detection unit of the BGO ECAL, an attenuation filter, an optical rubber, and a PMT with three dynodes' readouts are coupled on each end of the BGO.

Fig.2 shows an MDU in the BGO ECAL. The 600 mm long BGO crystal used is produced by Shanghai Institute of Ceramics, Chinese Academy of Sciences, which is the longest BGO single crystal so far [12]. An optical rubber and an attenuation filter are inserted between the PMT and the crystal. The optical rubber, with a good transparency of ~90% for 480nm wavelength photons, is used to cushion against mechanical stress to protect the fragile and hard PMT and crystal. The attenuation filter, an exposure film, is designed to tune the amplitudes of signals by attenuating the flux of scintillation lights injecting into the PMT with optional factors. In order to further extend the dynamic range, the lights generated in the crystals are read out unequally from the two ends by applying the attenuation filters with different factors in each MDU. Where, the less attenuated end is signed as "0 end" and the more attenuated end is signed as "1 end". By testing with cosmic rays, the MIP peaks measured from the two ends of the crystals are tuned to about 500 fC for 0 end and 100 fC for 1 end.

To construct the BGO ECAL detector, the MDUs were assembled into a honeycomb carbon fiber frame, in which there are $14 \times 22$ rectangle holes of 26mm $\times$ 26mm alternated in an orthogonal way. The distance between the centers of the adjacent layers is 29mm, and the pitch of the bars in the layers is 27.5mm.The gaps between the BGO crystals and the carbon fiber were filled with black silicon rubber for light blocking and mechanical steadiness [13]. Since the arrangement of the BGO bars and PMTs was very compact, the high voltage supports are provided together for 44 PMTs in every two layers of the four sides of the BGO ECAL.

The BGO ECAL provides the trigger signals from layers 1-4 and 11-14 for the DAMPE. Since the trigger threshold is same for bars a layer, the signals of a layer should be uniform [14]. Thanks to a high light yield of the BGO crystal, we can adjust the light generated by the BGO passing to the PMT by an attenuation filter. This way, we get better uniformity for each layer to satisfy the trigger requirement. At the same time, a ratio of 1:5 for the two ends of a MDU is set to extend the dynamic range.

The sensitive volume of the BGO ECAL is composed of 308 MDUs. In the following sections, the discussion on the calibration and reconstruction is based on the MDUs.

## 2.3 The ASIC

The PMTs chosen are from the R5610A-01 type made by Hamamatsu, which has ten stages of charge amplification [15].The VA32 ASIC, developed by IDE AS, Norway [16], was adopted for the front-end electronics of the ECAL, and the version used for the prototype of the DAMPE detector is VA32HDR14.2. It is a thirty-two-channel, low-power ASIC for charge measurement, with a dynamic range from -3 pC to +13 pC, 1.8 μs shaping time, and an intrinsic noise of about 0.5 fC RMS [17].

## 2.4 Temperature monitoring

*E-mail: ylzhang@ustc.edu.cn

The PMTs and BGO crystals are thermal-sensitive [18-19].The signals from the MDUs will decrease with temperature increases. The temperature coefficient is about -1.2% at 0 degrees centigrade for the BGO ECAL [20], so temperature monitoring is very important. There were sixty-four thermo-resistors (forty-eight of them on the crystals and sixteen on the FEE chips) attached in the BGO ECAL to monitor the temperature, so the signals from each MDU can be corrected with its temperature coefficient.

## 3. Calibration

In order to test and calibrate the detector, a long-term cosmic ray test in the laboratory and during the environmental experiments was carried out. Also, several beam tests at CERN-PS and CERN-SPS were performed. The basic parameters, including the pedestal, MIP response, and the ratios of dynode outputs, were calibrated with the cosmic rays and beam particles.

### 3.1 Pedestal

The pedestal is the reference voltage level at the signal input end of the electronic chains. Its fluctuation appears as a generalized Gaussian distribution and reflects the noise level of the electronics. There are 2016 electronic channels used in the calorimeter, and 1848 of them are connected to the dynodes of the PMTs for signal readout, while others are floated for monitoring the electronic chains themselves.Fig.3 shows the spectra of the pedestals, and Fig.4 shows the noise level, with sigma of the Gaussian function, distribution of all 2016 channels. The noises of 1848 channels are about 7.1 ADC channels, which corresponds to ~6 fC injected into the FEE chips. Further, the spare ones (which are floated) with lower input capacitance have a lower noise level as a consequence. This also provides a method for checking the cable connection of the signals, especially for the Dy2s, since they are almost unresponsive with the low energy (<10GeV [11, 21]) deposit. In addition, the pedestals will slightly drift with the temperature (<1fC per degree), so real-time calibration is necessary, as the temperature changes acutely.

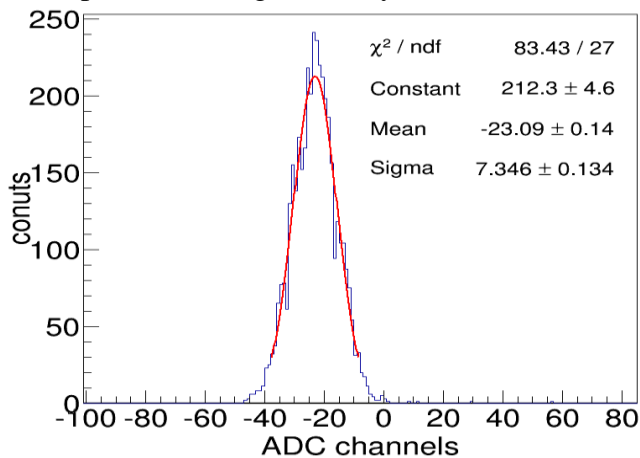

Fig.3.A typical pedestal spectrum of a single used channel, the red line is fitted by Gaussian function.

*E-mail: ylzhang@ustc.edu.cn

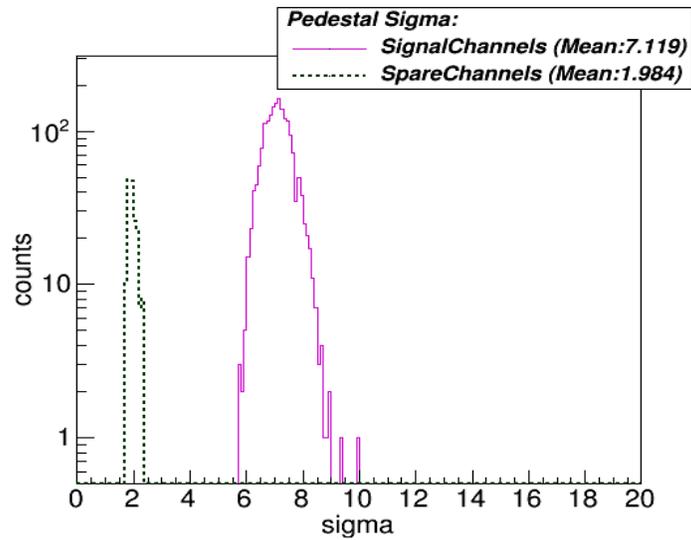

Fig.4.The pedestal widths (sigma) of all 2016 channels, the green dotted line is the sigma distribution of spare channels and the pink active line is of the signal channels

## 3.2 The ratios of dynode outputs

In the BGO ECAL, the PMTs coupled on both ends of the BGO bars are read out from three dynodes to achieve the high dynamic range for energy measurement. The ratios between the three dynodes are critical parameters for energy reconstruction. These parameters are extracted by fitting the scattering points with a linear formula using the least squares method.

In Fig.5, the slopes are the ratios of Dy2/Dy5 and Dy5/Dy8, which are about 0.02. This means that, with an energy deposit in the BGO bar, the signal of Dy8 is about 50 times that of Dy5 and 2500 times that of Dy2. Thus, only very high energy deposits can be used to calibrate the ratios of Dy2/Dy5. Which are corresponding to ~1500 MIPs for the 0 end and ~7500 MIPs for 1 end of the MDUs. The 243 GeV high-energy electron and 400GeV proton beams were used to scan the detector to calibrate the ratio. The electrons create showers in the front layers of the 14 layers and the protons tend to create showers in the later layers, so that the ratios of Dy2/Dy5 from all layers can be well calibrated.

The spread of fitted data points is caused by the noise fluctuation of the signals (~7.1 ADC, see the "Pedestal" section). When fitting the data with a linear function, a low range at 70 ADC (~10 times of pedestal sigma) is set for the lower signal dynodes (ordinate axes in Fig.5), which is corresponding to ~3500 ADC for the horizontal axes, a high range at 12000 (~10 pC, when higher than that, the saturation effect becomes significant) is set for the higher signal dynodes (horizontal axes in Fig.5).

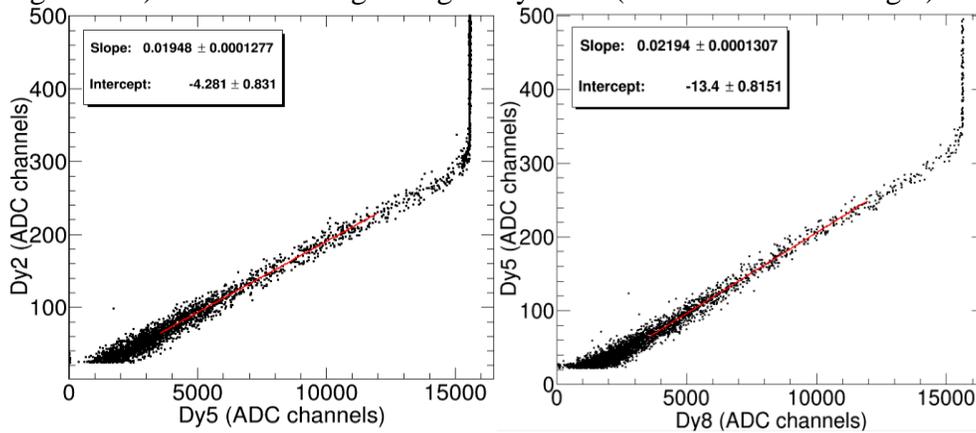

Fig.5. The typical dynode ratios of Dy2/Dy5 (left) and Dy5/Dy8 (right), the red line is the linear fitting function

*E-mail: ylzhang@ustc.edu.cn

## 3.3 Attenuation effect of BGO crystals

The fluorescent lights produced at the emission point are attenuated with an exponential function when transmitting along the BGO bars. The signals measured from both ends of the crystals are signed as $A_0$ (0 end) and $A_1$ (1 end). They can be expressed as:

$$A_0 = k_0 F_p e^{-x/\lambda} \quad (3.1)$$

$$A_1 = k_1 F_p e^{-(L-x)/\lambda} \quad (3.2)$$

where, $x$ is the distance from the position where the incident particle hits the crystal to 0 end of the bar. It is determined by the BGO ECAL's own tracks, which are obtained by fitting the fired crystals into the X-Z and Y-Z axes (see Fig.1, each involving 7 layers). $F_p$ is the primary fluorescent light yield produced in the crystals; $k_0$ and $k_1$ are the conversion coefficients between the lights at the ends of the bars, and the signals are recorded by the electronics, which include the filter, the rubber, and the PMTs etc. coupling efficiency between the optical elements, the quantum efficiency of the PMT, and the gain of the electronics, etc. $L$ is length of the crystals (600 mm), and $\lambda$ is the attenuation length.

According to Formulas 3.1-3.2, $A_0$ and $A_1$ change with the hit position of the particles. In order to achieve position-independent measurement of the energy deposit, a combined signal called $A_c$ is defined as:

$$A_c = \sqrt{A_0 A_1} = \sqrt{k_0 k_1} e^{-L/2\lambda} F_p = k_c F_p \quad (3.3)$$

which changes only with $F_p$. So that $A_c$ will be used in MIP calibration and energy reconstruction in the following sections. On the other hand, a relation between $A_0$, $A_1$ and $x$ can be obtained from Formula 3.1-3.2.

$$x = \frac{L - \ln(\frac{A_0 k_1}{A_1 k_0})\lambda}{2} \quad (3.4)$$

With this Formula (3.4), the $\lambda$ can be obtained by fitting the relation curve between $\ln(\frac{A_0 k_1}{A_1 k_0})$ and $x$. Then, it provides another way to obtain the hit positions ($x$) along the crystals by using the asymmetry between $A_0$ and $A_1$. The $k_0/k_1$ is determined by the ratios between the MIP peaks of both ends (when the cosmic rays that incident in the middle of bars are selected ($x = L/2$), the $k_0/k_1$ will equal to $A_0/A_1$).

With a long term (~72 hours) cosmic-rays test, the $\ln(\frac{A_0 k_1}{A_1 k_0})$ changing with the $x$ is plotted (Fig.6), the $\lambda$ is ~ 1550mm calculated by the fitting parameters.

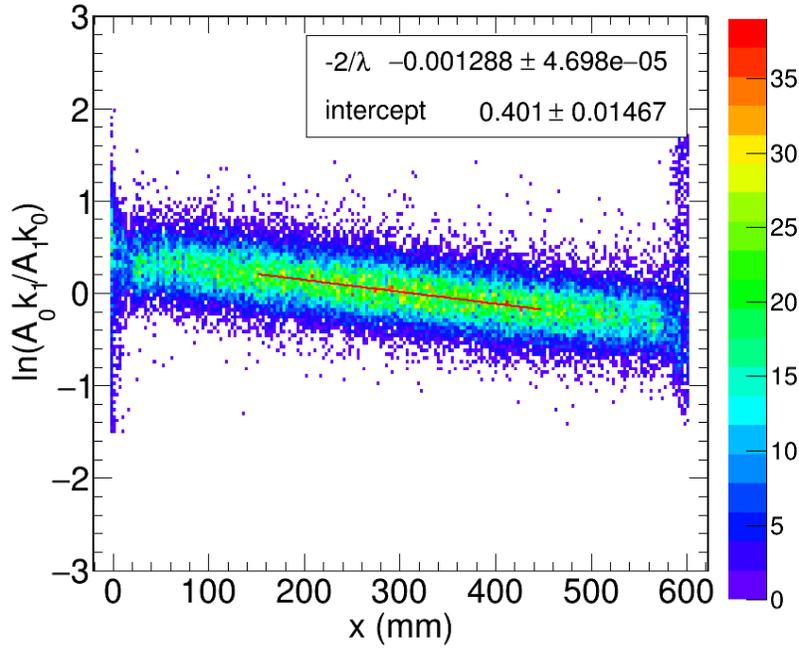

Fig.6. The scattering plot of $\ln(\frac{A_0 k_1}{A_1 k_0})$ versus $x$, the right palette shows the counts. A conservative fit range from 150 to 450 mm used to avoid the effect of the edges.

Also, the curves of $A_0$, $A_1$ and $A_c$ versus the $x$ are shown (Fig.7), $A_c$ is shown less dependent on the $x$ as expected.

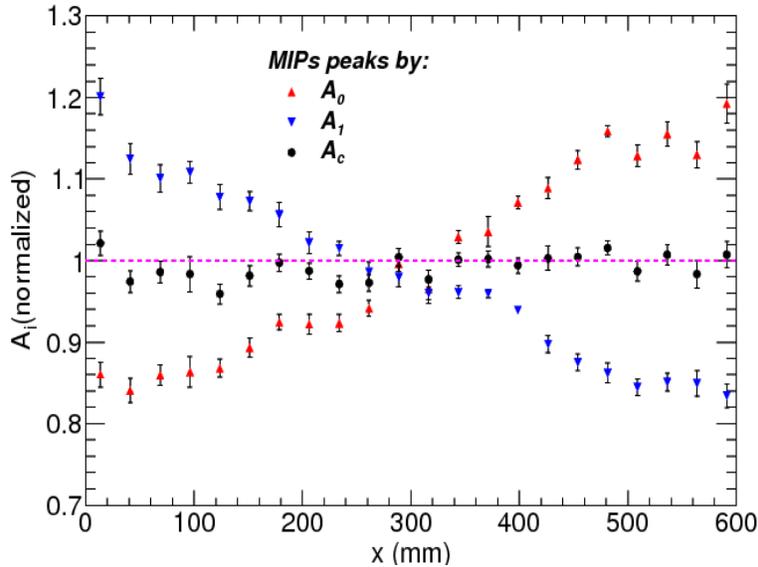

Fig.7. The MIP peaks of $A_0$, $A_1$ and $A_c$ change with the hit positions along the BGO crystal (tested with cosmic-rays).

## 3.4 MIP responses

The MIP energy deposit is the unique gauge that has been well studied. For the BGO ECAL, cosmic muons, beam hadrons and muons were used to calibrate the MIP responses on the ground level (i.e., protons in space for in-orbit calibration). A long-duration cosmic rays test was carried out to study the basic performance of the instrument and to calibrate the MIP responses. The typical MIP spectra from each end of the crystals are shown in Fig.8 and Fig.9 ($A_0$, $A_1$ and $A_c$ were defined in Section 3.3). The relative

*E-mail: ylzhang@ustc.edu.cn

"Width" compared with "MP" (the most probable value) is ~6%, , was fixed according the Geant4 based simulation when fitting the MIP spectra.

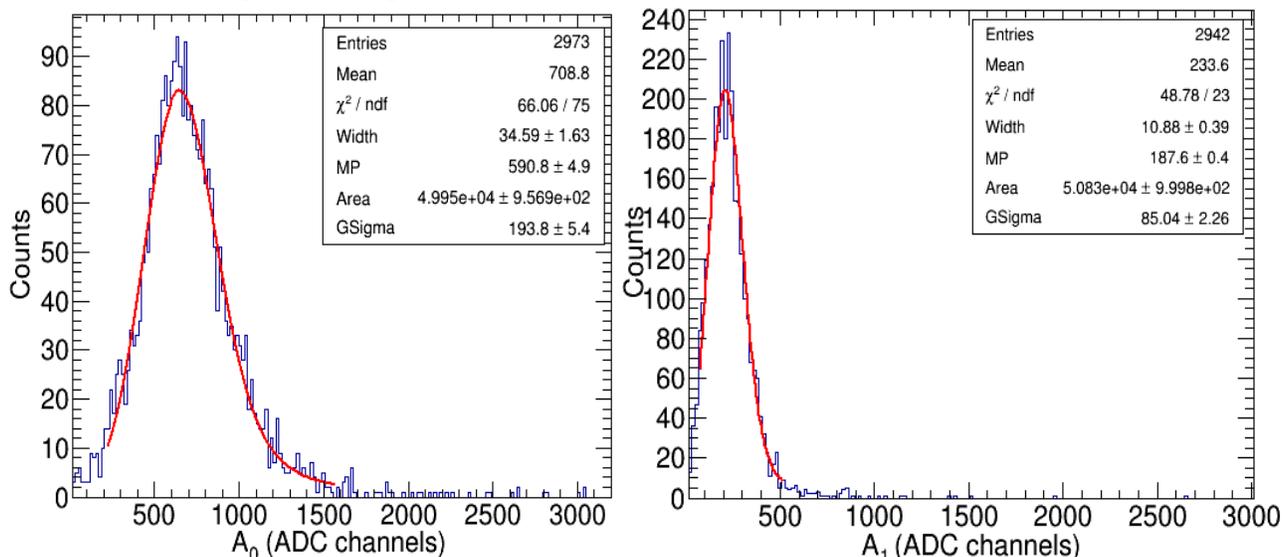

Fig.8. The cosmic rays MIP spectra of $A_0$ and $A_1$, the peaks are adjusted to ~ 500 fC (~600 ADC channels) for $A_0$ and ~ 100 fC (~120 ADC channels) for $A_1$

$A_c$ is less dependent on the hit positions, so that, the its MIP responses (Fig.9) are employed as the reference energy gauge when reconstructing the energy with the combined signals.

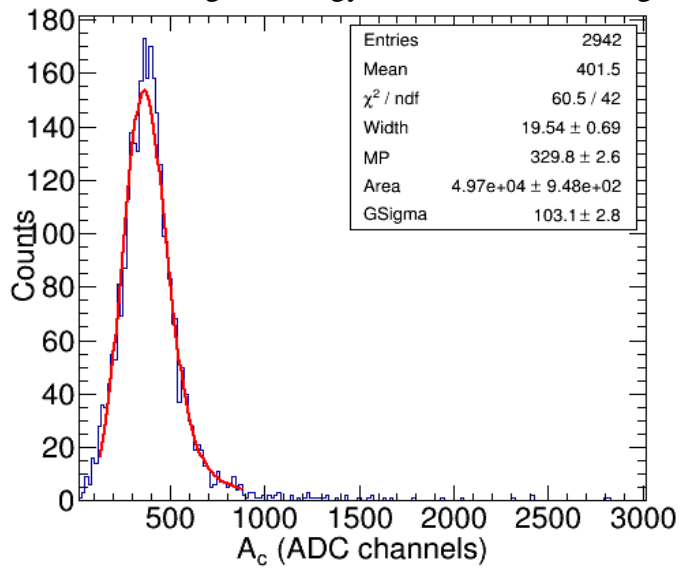

Fig.9. the MIP spectrum of $A_c$, the red line is the fitting function of Landau convoluted Gaussian.

## 4. The results of beam test

### 4.1 Test beams

The beam test was carried out at CERN-PS T9 and CERN-SPS H4 from October to November of 2014 using electronand proton beams in the energy region from 0.5 GeV to 243 GeV and from 10 GeV to 400 GeV, respectively, as well as 10 GeV π- and 150GeV muon beams for calibration. A beam test model with the same geometry and beam test setup was developed to simulate the MIPs and electrons by Geant4.

*E-mail: ylzhang@ustc.edu.cn

Otherwise, pedestals were monitored frequently (once an hour) with random triggers throughout the test duration.

## 4.2 Energy reconstruction

The energy reconstruction was processed by cutting the appropriate pedestals, choosing signals from the proper dynodes [11, 21-22], and normalizing them with the MIP peaks. The overall raw energy was taken to be the sum of the crystals' energies. Even though the BGO ECAL is a homogeneous electromagnetic calorimeter, some un-sensitive materials in the ECAL, such as the carbon fiber and the black rubber compose the dead areas. The incident particles also lose some energy in the frontier sub-detectors (PSD and STK) before the BGO ECAL. So that the raw energy $E_{CalRaw}$, that shown in Fig.10 and Fig.11, was slightly lower than the incident energy.

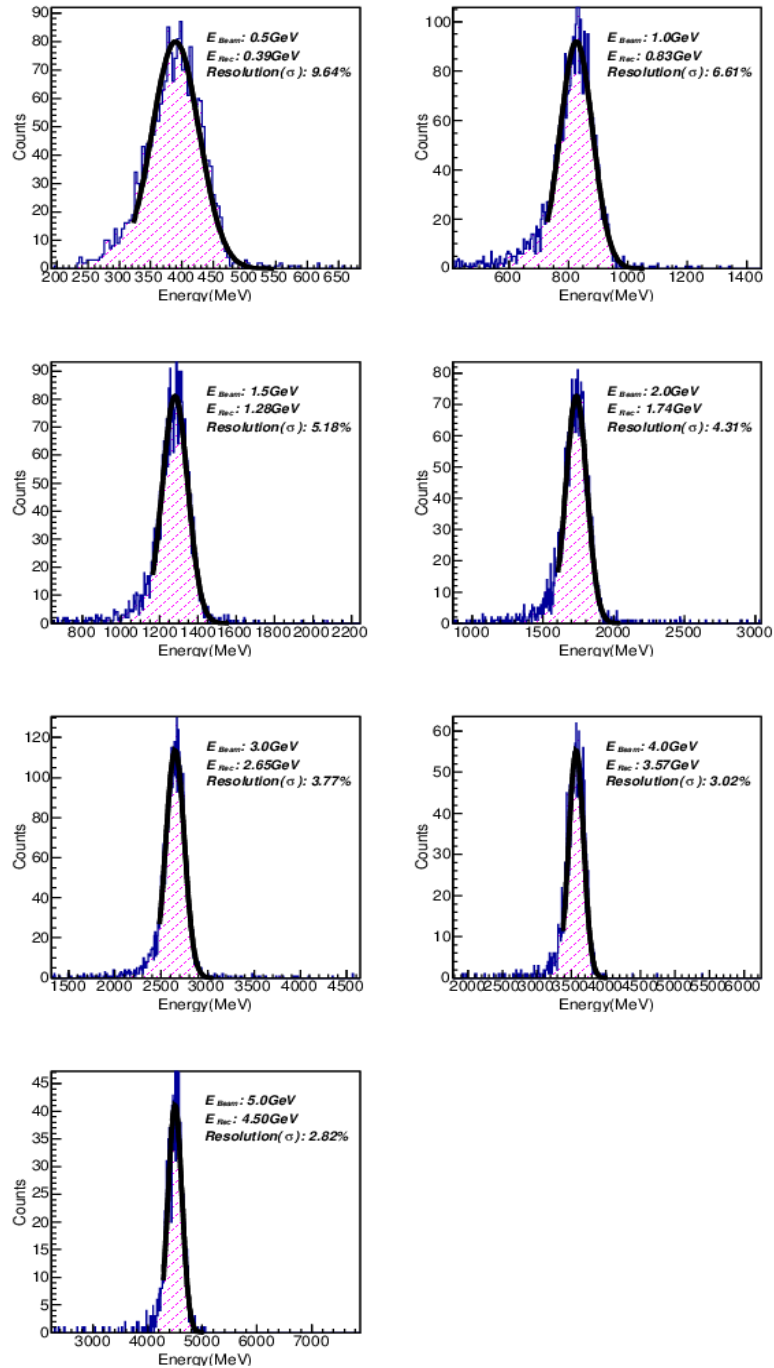

*E-mail: ylzhang@ustc.edu.cn

Fig.10. The reconstructed raw energy spectra of electrons from 0.5 to 5 GeV （at CERN-PS T9 beam line）

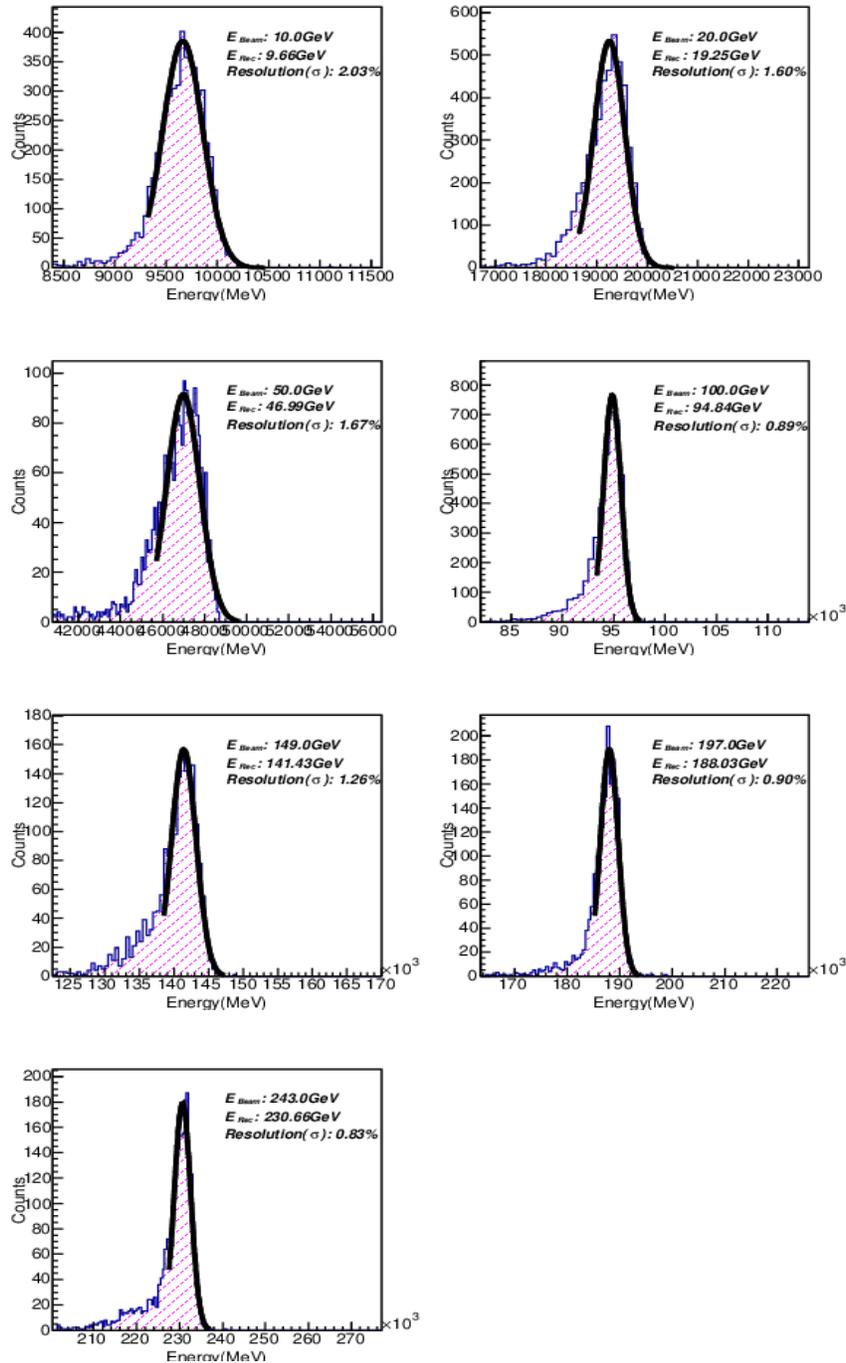

Fig.11. The reconstructed raw energy spectra of electrons from 10 to 243 GeV （at CERN-SPS H4 beam line）

In order to compare with the real data, some fluctuation derived from instruments should be considered in the Geant4-based simulation. The energy deposit in the crystals was readout and digitized with the calibrated parameters to consider the noise level of the electronics, the attenuation of the crystals, the statistic fluctuation, and the different level of reletive noise of the dynodes. Then, the overall raw energy deposit can be reconstructed as the way of the real data processed.

Fig.12 and Fig.13 show comparisons of the energy and energy resolution between the beam test data and the simulation data. The good agreement between the simulation and real data confirms the validity of the calibration and reconstruction. The energy resolution at 800 GeV of about 0.8% can be calculated with the parameters shown in Fig.13. The Geant4 simulation shows that the energy leakage from the bottom of the

*E-mail: ylzhang@ustc.edu.cn

BGO ECAL is not significant for electron energy below 1 TeV. Additionally, some methods developed with the Geant4 simulation could be applied to the real data to correct the raw energy to the incident energy [23].

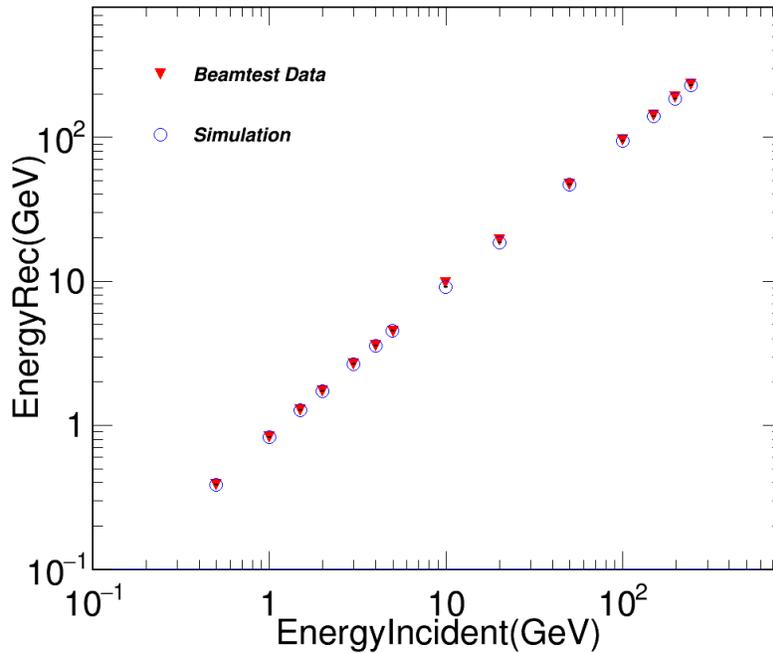

Fig.12. There construction raw energies versus the incident electron energies

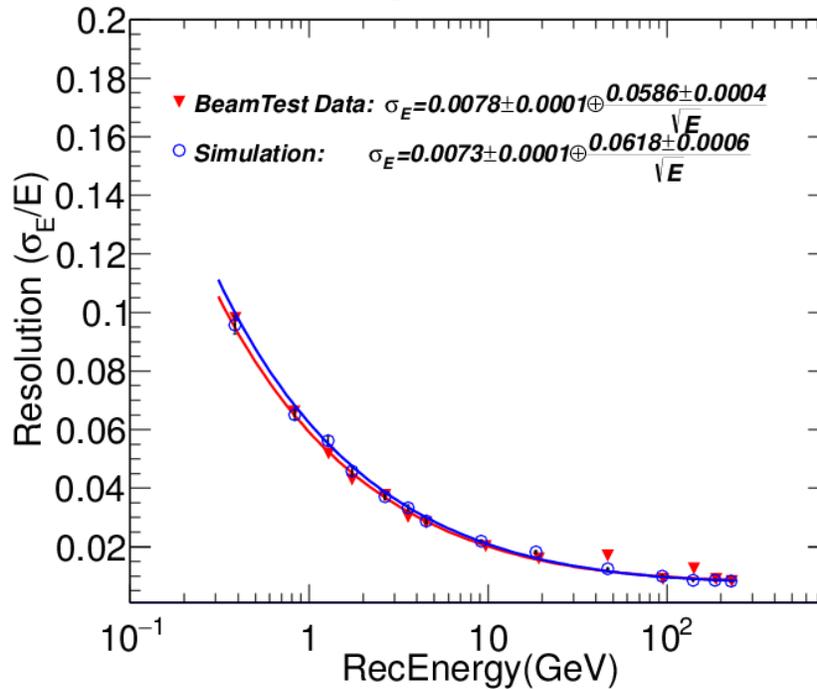

Fig.13. The energy resolutions change with the incident electron energies.

## 5. Conclusion

The DAMPE engineering model had been built and the systemically tested in the environmental experiments and calibrated with the cosmic rays and test beam. Basic methods and results of calibration were discussed. The good energy resolution of better that 1% at 100GeV is obtained by raw energy reconstruction. The agreement of Monte Carlo and real data on electron energy reconstruction confirmed the validity of the calibration and reconstruction. Also, a study of energy correction for the electron beam data were performed and published in a separate paper [23].

*E-mail: ylzhang@ustc.edu.cn

Further studies, like ions beam test, e/p identification study with proton beams at CERN were also carried out during the last several months, the data analysis works is under progressing.

# Acknowledgments

The authors wish to thank professor Xin Wu from University of Geneva), professor Giovanni Ambrosi from INFN Perugia and professor Mario Nicola Mazziotta from INFN Bari for their crucial contributions on the beam application and test setup. This work was supported by the Chinese 973 Program, Grant No. 2010CB 833002, the Strategic Priority Research Program on Space Science of the Chinese Academy of Science, Grant No. XDA04040202-4.

*E-mail: ylzhang@ustc.edu.cn